\documentclass[reprint,
 amsmath,amssymb,
 aps,
 pre,
]{revtex4-1}
\usepackage{graphicx}% Include figure files
\usepackage{dcolumn}% Align table columns on decimal point
\usepackage{bm}% bold math
\usepackage{float}
\usepackage{xcolor}
\usepackage{subfigure}
\usepackage[abs]{overpic}
\usepackage{hyperref}
\usepackage[dvipsnames]{xcolor}
\usepackage{lineno}
\usepackage{comment}

\usepackage{soul} %To stike through a text

\begin{document}
\makeatletter
\newcommand*{\rom}[1]{\expandafter\@slowromancap\romannumeral #1@}
\makeatother
\preprint{APS/123-QED}
%=======================

\title{Geometry-induced orientational trapping dictates escape of active rods in corrugated channels}

%\title{Topology-induced orientational trapping dictates escape of active rods in corrugated channels}

\author{Baburao Simma}
\affiliation{Department of Physics, National ~Institute ~of ~Technology, 
 Warangal, 506004, India}

\author{Rahul Bag}
\affiliation{Department of Physics, National ~Institute ~of ~Technology, 
Warangal, 506004, India}

\author{Hari K. Reddy }
\affiliation{Department of Physics, National ~Institute ~of ~Technology, 
Warangal, 506004, India}

\author{Radha Ramana}
\affiliation{Department of Physics, National ~Institute ~of ~Technology, 
Warangal, 506004, India}

\author{Vrinda Garg}
\affiliation{Department of Physics, National ~Institute ~of ~Technology, 
Warangal, 506004, India}

\author{Surya K. Ghosh}
\email{skghosh@nitw.ac.in}
\affiliation{Department of Physics, National ~Institute ~of ~Technology, 
Warangal, 506004, India}

%\date{\today}

\begin{abstract}

The non-equilibrium transport of anisotropic active matter through geometrically confined environments is fundamentally governed by the interplay between self-propulsion and geometry-induced entropic barriers. Using overdamped Langevin dynamics, we reveal contrasting escape mechanisms of self-driven rods in narrow corrugated channels by comparing smoothly varying sinusoidal (SC) geometries to circular (CC) cavities with localized, uniform curvature. We observe a strictly non-monotonic dependence of the mean first escape time (MFET) with increasing Péclet number ($Pe$) in the CC channel, contrary to the monotonic dependence in the SC channel. This non-monotonicity emerges from a competition between activity-enhanced escape at low-to-intermediate propulsion and persistent, geometry-induced orientational trapping along the circular boundary at high propulsion, which inhibits reorientation toward the exit neck. This regime crossover is further corroborated by trajectory-to-trajectory escape fluctuations and bimodal radial probability distributions reflecting persistent wall-collision and wall-aligned sliding transport. The self part of van-Hove correlation function exhibits distinct anisotropic non-Gaussian displacements in the SC channel, in contrast to symmetric profiles with extended active-trapping plateaus in the CC cavities. These results establish local channel topology as a critical determinant of confined active transport, with direct implications for microfluidic sorting, biological navigation, and targeted drug delivery in corrugated environments.

\end{abstract}

\maketitle

%\tableofcontents

\section{Introduction}

The non-equilibrium active matter systems are composed of self-driven units capable of converting local energy into persistent motion\cite{Elgeti_2015, Bechinger2016, Ramaswamy_2017, Hartmut_JCP_2020, romanczuk2012active}. Examples of such non-equilibrium particles are omnipresent in all domains, ranging from biological microswimmers like \textit{Escherichia coli} and spermatozoa \cite{chen_PhysRevLett_bacteriaClusters, Sokolov_PhysRevLett_SwimmingBacteria, Dombrowski_PhysRevLett_BacterialDynamics, Ishikawa_PhysRevLett_2011, Rousselet_PhysRevLett_2011, J.Schwarz_PNAS_2012, Ingmar_Science_2005, VasilyKantsler_PNAS_2013} to synthetic bimetallic (Au-Pt) nanorods \cite{Paxton_JACS_2004}. Among these, active nanorods represent a fundamentally important class of anisotropic particles. Their shape anisotropy introduces a strong coupling between translational and rotational degrees of freedom \cite{Fazli_JCP_2016}, a feature that is particularly dominant in confined geometries, where topological constraints of the boundary dictate the transport and escape kinetics of elongated particles.

The transport through confined geometries is a central problem in physical chemistry and biophysics, relevant to molecular transport in porous membranes, ion conduction, microfluidic devices and drug delivery \cite{Reguera2001, Burada2009, Howse2007, roy2023overcoming, Maity_2026,  Kaiser2012, Fily2014, kim2022}. At the microscale, confinement alters particle motion by coupling geometric constraints with thermal fluctuations and interaction forces, often triggering nontrivial transport behaviors mechanism. While passive Brownian particles overcome these barriers via classical Kramer's-type thermal activation, where escape rate scales as $k \propto exp(-\Delta U/k_BT)$ \cite{Martin_Lisal_SoftMatter_2025, Lee_2013}, active particles exhibit fundamentally different escape phenomenology\cite{Caprini_JCP_2021}.

The escape and transport through confined channels, porous media, and gel-like environments have been studied extensively over the past decade due to their significance in intracellular transport, filtration, and microfluidic applications. Initial studies focused on Active brownian particles (ABP), demonstrating that activity noticeably elevates the escape from confined cavities, barrier crossing, corrugated channels, and entropic landscapes compared with passive transport\cite{Ao_EPJ_2014, Woillez_PRL_2019, TanwiDebnath_PKGhosh_JCP_2021, Nayak2023, garg2024crowding, Olsen_PRR_2020}. Subsequently, the chiral ABPs have gained attention because their transport and escape dynamics are fundamentally different from those of ABPs due to the intrinsic angular movement that traces circular trajectories that can enhance or suppress the transport mechanism, depending on channel topology and chirality strength\cite{Ao_EPL_2015, Upadhyaya2024, Iyaniwura_SoftMatter_2026}. In recent times, the escape and transport of active polymers and tangentially driven filaments through narrow pores and confined geometries have emerged as a prominent research direction, where activity, chain flexibility, stiffness, and excluded-volume effects mutually control conformational dynamics \cite{KR_Prathyusha_PRL_2026, Elgeti_Gompper_SoftMatter_2018, MingmingDing_SoftMatter_2023, MingmingDing_SoftMatter_2024, Bhowmik_Ramaswamy_PRE_2026}. Laterally, active dumbbells, Janus dimers, and active rods have been studied extensively as minimal models of deformable micro-robots, revealing that internal degrees of freedom and anisotropic propulsion produce rich escape phenomena, including wall accumulation and enhanced diffusivity \cite{Fazli_JCP_2016, kumar2023, Tiwari_SPSingh_SoftMatter_2024}.

Corrugated channels characterized by periodically modulated boundaries serve as an idealized model for porous media and biological systems. In such geometries, the spatially varying confinement induces localized entropic traps \cite{Reguera2006, Agosta_Mikhail_JCP_2024, ghosh2012chemosensing}. For active matter, these geometric corrugations strongly combine with the particle's persistence length and propulsion strength, resulting in rich, counterintuitive transport phenomena \cite{DebasishMondal_PhysRevE_2010, garg2026semiflexibility}. Extensive investigations of passive transport in corrugated channels have elucidated phenomena such as geometry-induced drift and mobility suppression \cite{Ghosh_Schmid_2012, Nayak2023, pattanayak2019enhanced, Dey2018}. 
Extensions to active particles have revealed enhanced transport, affinity to the wall boundary \cite{Wensink_PhyRevE_2008, KangChen_SoftMatter_2020, Tiwari_SPSingh_SoftMatter_2024}, accumulation near bottlenecks \cite{SilkeHenkes_Marchetti_PRE_2011}, and activity-controlled escape rates \cite{MingmingDing_SoftMatter_2023, MingmingDing_SoftMatter_2024, Upadhyaya2024, Chen2022}. Mean escape times decrease monotonically for both rod-like particles and active-head polymers exiting circular cavities through narrow openings\cite{kumar2023, Sahoo2025}.

However, the majority of these studies focus on generic confinements or isotropic particles\cite{angelani_EPJE_2024, Caprini_JCP_2019}. A systematic investigation into how different corrugation \textit{morphologies} govern the orientation-dependent escape dynamics of rigid active nanorods remains limited. Here, we address this gap by directly comparing the non-equilibrium escape dynamics of an active nanorod confined in two different topologies: a sinusoidal corrugated (SC) channel and a circular corrugated (CC) channel. The two geometries have almost equal confinement size, but they differ fundamentally in their blob curvature. The SC channel has a smoothly varying curvature, enabling continuous reorientation and alignment-assisted transport. On the other hand, the CC channel generates localized regions of strong, abrupt curvature change between the blob and the neck, which result in deep entropic traps and severely constrain the rotational freedom, thereby restricting the rod from finding an escape route.

This study has relevance to the structural complexity of real-world biological and synthetic microenvironments. In physiological systems, rod-like pathogens such as \textit{Helicobacter pylori} must actively navigate through the viscoelastic, sterically constrained mucin polymer networks of the gastric mucosa, which naturally present periodically varying geometric bottlenecks \cite{Celli_PNAS_2009, Walker_AIP_2023}. Similarly, chemical engineering and directed drug delivery applications \cite{roy2021kinetics, garg2025interplay}, active catalytic nanomotors such as asymmetric mesoporous silica nanoparticles designed for targeted drug delivery, must pass through porous matrices and biological barriers \cite{Ma_JACS_2015, Wang_ChemRev_2015}. In microfluidic technologies, sorting platforms utilizing deterministic lateral displacement (DLD) and asymmetric ratchets rely entirely on specific local boundary curvatures to filter cells and active nanocarriers \cite{McGrath_DLD_2014, Bechinger2016}.

In all these scenarios, active transport efficiency is not governed merely by the average pore size, but by the precise topography of the local curvature. It remains an open question whether confining walls guide an anisotropic particle smoothly via alignment (analogous to our SC channel) or induce deep entropic trapping within sharp, localized cavities (analogous to our CC channel). Understanding how these geometric differences regulate intermittent trapping and escape is therefore crucial for optimizing microfluidic sorting technologies and predicting the invasive transport dynamics of active biological agents.

Our study highlights the effect of channel anisotropy on the escape dynamics of rod-like particles. In particular, the escape time has shown different behavior in two channels considered. MFET shows monotonic behavior for SC channel, while it shows non monotonic trend in CC channel. We notice bimodal patterns in the radial probability distribution function (RPDF) and anisotropic effect of channel on self part of van-Hove correlation function (VHCF). Our analysis clearly establish the difference of escape dynamics in two distinct topological channels, including effect of rotational degree of freedom.

The paper is organized as follows: Section II details the theoretical model and simulation methodology. Section III presents the main results, contrasting the spatial organization, macroscopic escape statistics, and microscopic displacement correlations between the two channel geometries. Finally, Section IV summarizes our conclusions and outlines possible extensions of this work.

%=========================================================================
\begin{figure*}
    \includegraphics[width=16cm]{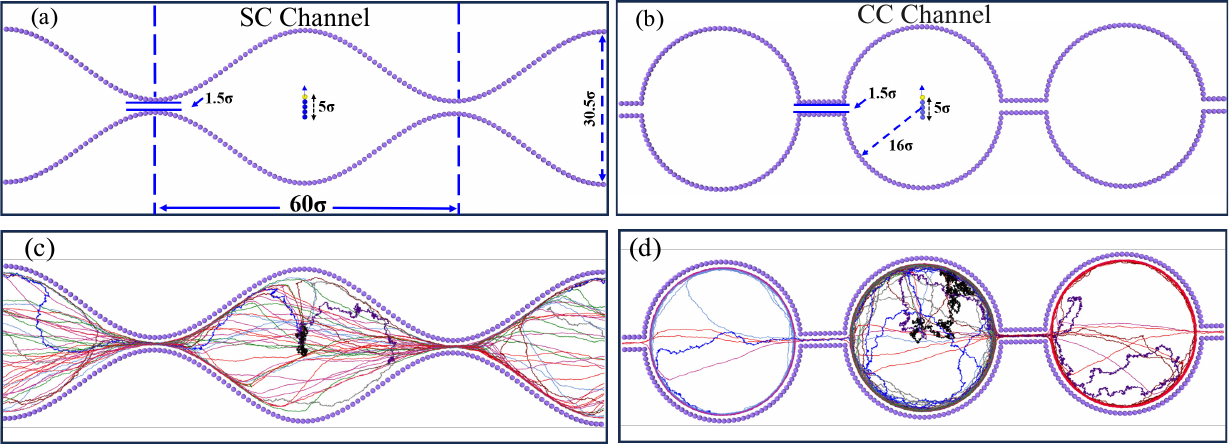}
    \caption{Schematic of the confinement geometries and representative active rod trajectories. (a) The sinusoidal corrugated (SC) channel, characterized by smoothly varying curvature. (b) The circular corrugated (CC) channel, featuring highly localized circular cavities. In both cases, the rigid active rod of length $5\sigma$ is depicted in its unbiased initial vertical orientation at the center of the channel blob. The trajectories of the center of mass of the active rod within the (c) SC channel and (d) CC channel. Different colors corresponds to the distinct values of P\'{e}clet number, $Pe = 0$ (black), $1$ (indigo), $3$ (blue), $5$ (gray), $10$ (maroon), $20$ (cornflower blue), $30$ (green), $40$ (red), and $60$ (violet red) respectively . In our model, periodic boundary conditions are applied along the longitudinal $x$-axis, and the channel walls are treated as purely repulsive rigid boundaries that permit alignment-assisted sliding. The width of the channel openning is set to $1.5\sigma$, here $\sigma$ the elementary unit of length.}
  \label{Model}
\end{figure*}
%=========================================================================

\section{Model and Methods}
\subsection{Particle Models and Channel Geometry}

We model the active particle as a rigid rod composed of $5$ linearly fused beads, each with mass $m$ and diameter $\sigma$. The total length of the rod is thus $l = 5\sigma$. The rod is treated as a single rigid body possessing both translational and rotational degrees of freedom; see Fig. \ref{Active_Rod_Trj}(a). The system is described using standard Lennard-Jones (LJ) reduced units, where length, energy, and mass are expressed in units of $\sigma$, $\varepsilon$, and $m$, respectively. Consequently, the characteristic timescale is $\tau = \sigma\sqrt{m/\varepsilon}$, and the thermal energy is set to $k_BT = 1.0\varepsilon$.

To investigate the role of topological confinement on escape dynamics, we restrict the rod's motion to two distinct periodic two-dimensional geometries, illustrated in Fig. \ref{Model}. The first is a sinusoidal corrugated (SC) channel Fig. \ref{Model}(a), characterized by a smoothly varying, anisotropic curvature, with the channel architecture function is given by $S_{\pm}(x) = \pm \frac{1}{2}[w + (y_{m}-w)sin(\pi x/ x_m)]$\cite{Burada2009, Nayak_PK_Ghosh_JPCB_2026, Biswas_Kudrolli_2023}. Here $w$, defines the width of the horizontal opening. The second is a circular corrugated (CC) channel Fig. \ref{Model}(c), featuring isotropic, highly localized curvature. The circular channel accomplished by circular cavity of radius $R$ connected with linear horizontal tube of fixed length ($l_t = 4.0 \sigma$). The circular cavity has two horizontally opposite openings \cite{Biswas_Kudrolli_2023}. In both geometries, the channels are constructed as rigid, and static boundaries. The transverse confinement is defined by a maximum half-width of $y_{m} = 15\sigma$ for the SC channel and a circular blob radius of $R = 16\sigma$ for the CC channel, with both geometries converging to a narrow escape neck of width $w = 1.5\sigma$ \cite{Burada2009, Ao_EPJ_2014, Biswas_Kudrolli_2023}. Although sharing comparable confinement scales, the SC and CC channels diverge in boundary symmetry and axial periodicity, driving distinct localized transport dynamics.

\subsection{Potentials and Equations of Motion}

The excluded volume interaction between the constituent beads of the active rod and the channel walls is modeled using the purely repulsive Weeks–Chandler–Andersen (WCA) potential,

%==========================================
\begin{equation}\label{eq:WCA}
U_{\mathrm{WCA}}(r_{ij})=
\begin{cases}
4\varepsilon_{ij}
\left[
\left(\dfrac{\sigma_{ij}}{r_{ij}}\right)^{12}
-
\left(\dfrac{\sigma_{ij}}{r_{ij}}\right)^{6}
\right]
+\varepsilon_{ij},
& r_{ij} \le r_{cut}, \\[10pt]
0,
& r_{ij} > r_{cut} .
\end{cases}
\end{equation}
%==================================================

where $r_{ij}$ is the distance between an interacting pair, $\sigma_{ij}=\frac{\sigma_i + \sigma_j}{2}$ is the effective interaction diameter, and the cutoff distance is $r_{cut} = 2^{1/6}\sigma_{ij}$. The interaction strength is uniformly set to $\varepsilon_{ij} = 1.0 \varepsilon$.

The translational stochastic dynamics of the self-driven rod are initially formulated using the underdamped Langevin equation,

\begin{equation} \label{Langevineq_UD}
m\frac{d^{2}{\boldsymbol{r}}(t)}{dt^2} = -\nabla U_{tot} - \gamma_t \frac{d\boldsymbol{r}}{dt} +\boldsymbol{\xi}(t) + F_a \boldsymbol{e},
\end{equation}

where $\boldsymbol{r}(t)$ is the center of mass (COM) position, $U_{tot}$ is the total potential energy exerted by the boundaries, and $\gamma_t = m/\tau_d$ is the translational friction coefficient. The term $\boldsymbol{\xi}(t)$ represents Gaussian white noise satisfying the fluctuation-dissipation theorem, $\langle \xi (t) \rangle = 0$, $\langle\xi_i(t)\cdot\xi_j(t')\rangle = 4\gamma_t k_BT\delta_{ij}\delta(t-t')$, where $k_B$ is the Boltzmann constant and $T$ is the constant system temperature. The active self-propulsion force, $F_a$, acts continuously along the rod's main axis, defined by the unit orientation vector $\boldsymbol{e}$.

To model the physically relevant low-Reynolds-number regime, we take the overdamped limit, wherein the inertial term has no significant effect. To ensure practically overdamped regime, we chosen high value of $\gamma_t = 500$ \cite{Gompher_Winkler_JCP_2021}, that results short momentum relaxation time $\tau_m = \frac{m}{\gamma_t} \approx 2 \times 10^{-3}$(in units of Lennard-Jones the time is measured as $t_0 = \sqrt{\frac{m\sigma^2}{\varepsilon}}$, $\sigma$, m, and $\varepsilon$ are the units of length, mass, and energy). On the other hand, the Brownian timescale defined as the time it takes the particle to diffuse a distance of it's own size is given by $\tau_B = \frac{\sigma^2}{D_t} \approx 500$, here $D_t= \frac{k_BT}{\gamma_t}$ is the translational diffusion constant.  The high value of friction coefficient, we obtain $\tau_m = 2 \times 10^{-3} << \tau_B = 500 t_0$, which ensures the negligible inertial effects and this approach widely used in literature \cite{Gompher_Winkler_JCP_2021}. This reduces Eq. \ref{Langevineq_UD} to the overdamped translational equation

%======================================
\begin{equation} \label{Langevineq_OD}
\gamma_t \frac{d\boldsymbol{r}}{dt}= -\nabla U_{tot} +\boldsymbol{\xi}(t) + F_a \boldsymbol{e}.
\end{equation}
%======================================

From this formulation the velocity auto correlation function decays exponentially
%=====================================
\begin{equation} \label{VACF_equation}
\langle v(t) v(0) \rangle = v_0^2 e^{-D_rt}, t>0, 
\end{equation}
%====================================

where $D_r = 1/\tau_r$ is the rotational diffusion rate, and $v_0 = F_a/\gamma_t$ is the self-propulsion speed in a direction denoted by unit vector $\boldsymbol{e}$. Consequently, in free space ($U_{tot} = 0$), the mean squared displacement (MSD) exhibits the following exact analytical expression

\begin{equation} \label{MSD_freespac}
\langle \Delta r^2 (t)\rangle = 4D_tt + \frac{2v_0^2}{D_r^2} (D_r t + e^{-D_rt} - 1),
\end{equation}

where $D_t$ and $D_r$ are the translational and rotational diffusion coefficients, respectively.
The orientational dynamics of the rod are similarly described by the rotational underdamped Langevin equation
%===============================================
\begin{equation} \label{rotational_Langevineq_UD}
 I\frac{d^{2}\theta(t)}{dt^2} = - \gamma_r \frac{d \theta(t)}{dt} + \eta(t),
\end{equation}
%=============================================

Applying the overdamped limit ($I \to 0$) yields the first-order rotational equation,
%================================================
\begin{equation} \label{rotational_Langevineq_OD}
\frac{d \theta(t)}{dt} = \frac{1}{\gamma_r} \eta(t).
\end{equation}
%================================================
 
Here, $\eta(t)$ is the Gaussian-distributed random noise with zero mean, variance $\langle\eta_\alpha(t) \eta_\beta(t')\rangle = 2D_r\delta_{\alpha \beta}\delta(t-t')$, and $\theta(t)$ is the instantaneous orientation angle of the rod, and the orientation unit vector is $\boldsymbol{e} = (\cos\theta(t), \sin\theta(t), 0)$. In two dimensions, rotational motion is generated by the out-of-plane stochastic angular torque, $\eta_z(t)$. The temporal evolution of the unit vector components is therefore given by

\begin{equation} \label{unit_vector_x_y}
\frac{d\boldsymbol{e_x}}{dt} = -\eta_z(t) sin(\theta(t)), 
 \frac{d\boldsymbol{e_y}}{dt} = \eta_z(t) cos(\theta(t)),
\end{equation}

%\begin{equation} \label{unit_vector_y}
%\frac{d\boldsymbol{e_y}}{dt} = \eta_z(t) cos(\theta(t))
%\end{equation}

From relation \ref{unit_vector_x_y}, the rotational change in the unit vector can be written as

\begin{equation} \label{rotaional_change}
\frac{d\theta}{dt} = \eta_z(t),
\end{equation}

From the above relations, it is evident that the autocorrelation of rotational vector $\langle e(t)e(0)\rangle = e^{-D_rt}$, here $D_r$ is the rotational diffusion constant which related to translational diffusion constant as $D_r = \frac{3 D_t}{\sigma^2}$, $\sigma$ is the size of the particle. To quantify the self propulsion we define dimensionless parameter P\'{e}clet number defined as the ratio of active advection to thermal diffusion \cite{kim2022},
%=======================================================
\begin{equation} \label{Peclet_number}
 Pe  = \frac{v_0\sigma}{D_{t}}=\frac{F_a\sigma}{k_BT}.
\end{equation}
%========================================================
In our study, $Pe$ serves as the primary control parameter governing the active strength of the rod.

%\subsection{Simulation Details}

All simulations were performed using a Langevin thermostat, and the equations of motion were integrated using the velocity-Verlet algorithm in the LAMMPS molecular dynamics package \cite{PLIMPTON19951}. To ensure the escape orientation was unbiased, the rod was initialized at the exact center of the channel with a vertical orientation. The system was allowed an initial equilibration period of $1 \times 10^4$ time steps. Total simulation run for $3 \times 10^8$ time steps, with the COM coordinates recorded every $100$ steps for statistical averaging. In all simulations, energy is measured in units of $\varepsilon = k_BT$, m is the units of mass, and $\sigma$ is the unit of length, respectively, and the unit of time is $t_0 = \sqrt{\frac{m\sigma^2}{\varepsilon}}$. We fix temperature as $k_BT = 1.0$, and timestep $\Delta t= 5 \times 10^{-4}$. Hydrodynamic interactions are disregarded for simplicity; see, Table. C1 and C2 in supplementary for acronyms and simulation details respectively.

\subsection{Observables}

To characterize the transport properties \cite{garg2025protocols} and escape dynamics of the active rod, we first evaluate the mean squared displacement (MSD) of its center of mass (COM) \cite{SAXTON19971744, Ernst_2013, KUSUMI19932021}. The time-averaged MSD for a single trajectory over a total observation time $T$ and lag time $\Delta$ is defined as:

%====================================================
\begin{equation} \label{MSD_eq}
\overline {\delta^2(\Delta)} = \frac{1}{T-\Delta} \int_{0}^{T-\Delta} [r(t+\Delta) - r(t)]^2 \,dt,
\end{equation}
%====================================================

The MSDs are ensemble averaged over $N$ independent trajectories, $\langle \overline {\delta^2(\Delta)} \rangle = \frac{1}{N} \sum_{i=1}^{N} \overline {\delta_i^2(\Delta)}$. The MSD generally exhibits a power-law scaling, $\langle \overline {\delta^2(\Delta)} \rangle = 4D \Delta^\alpha$, where $\alpha = 1$ denotes normal Fickian diffusion and $\alpha > 1$ signifies the super-diffusive behavior characteristic of active motion at intermediate timescales.

To quantify the spatial organization, interaction of the active rod with the channel boundaries, and the escape characteristics, we numerically evaluated the radial probability density function $P(r)$ (RPDF) of center of mass (COM)\cite{kumar2023, Sahoo2025}. We define distance $r$ from the origin in two dimensional cartesian coordinate system. The probability distribution of distance ($r$) is calculated up until the rod successfully escapes, which is defined as the COM traversing the narrow neck of width $1.5\sigma$; see Fig. \ref{Model}. This definition represents the global radial distance and does not account for the spatial variation of the local wall position in the SC channel.

The escape kinetics are primarily characterized by the mean first escape time (MFET) \cite{Benichou2008, Grebenkov_2019, Chevalier_2011, Chen2022, Biswas2020, Biswas_Kudrolli_2023, Condamin2005, Mejia-Monasterio_2011, Mattos2012, kumar2023}. The MFET, $\langle \tau_{esc} \rangle$, is obtained by integrating over the probability density function of the first escape times, $P(\tau_{esc})$

\begin{equation} \label{MFET_equation}
\langle \tau_{esc} \rangle = \int_{0}^{\infty} \tau_{esc} P(\tau_{esc}) d\tau_{esc}.
\end{equation}

Here, $P(\tau_{esc})$ is the distribution function of the first escape time, and integration is taken over all the escape events. The first escape time is defined as the time taken by the rod’s center of mass (COM) to traverse the cavity and reach the escape boundary. For the sinusoidally corrugated (SC) channel, an escape event is recorded when the COM crosses either of the two blue dashed lines marking the cavity boundaries, as shown in Fig. \ref{Model}(a). For the circularly corrugated (CC) channel, escape is defined by the COM crossing the corresponding circular boundary; Fig. \ref{Model}(b). This criterion provides a consistent basis for determining the first-escape time in both channel geometries.

Further, to measure the relative statistical fluctuations in the escape dynamics, we evaluate the coefficient of variation (CV) 

\begin{equation} \label{CV_equation}
CV = \frac{\sqrt{\langle\tau^2_{esc}\rangle-\langle\tau_{esc}\rangle^2}}{\langle \tau_{esc} \rangle},
\end{equation}

Here, $\langle \tau_{esc} \rangle$ and $\langle \tau^2_{esc} \rangle$ are the first and second moments of FETs respectively. Although the description of coefficient of variation is extensive, it represents simply a number, which leaves the explanation is still unclear. The trajectory-to-trajectory variation in the first escape times can be quantified using the uniformity index (UI) effectively. The uniformity index ($\omega$) can be defined as\cite{KoushikGoswami_RMetzler_2024}:

\begin{equation} \label{UI_equation}
\omega = \frac{\tau^{'}_{esc}}{\tau^{'}_{esc} + \tau^{''}_{esc}}.
\end{equation}

Here, $\tau^{'}_{esc}$ and $\tau^{''}_{esc}$ are set of two random trajectories. From the definition of $\omega$,
if, $\tau^{'}_{esc} >> \tau^{''}_{esc}$, then $\omega \approx \frac{\tau^{'}_{esc}}{\tau^{'}_{esc}} \approx 1$; in such cases, we can observe a maximum peak at $w=1$ in the $P(w)$ vs $w$ plot, which corresponds to a large variation in the first escape time between trajectories.
Subsequently, $\tau^{'}_{esc} << \tau^{''}_{esc}$, then $\omega \approx \frac{\tau^{'}_{esc}}{\tau^{''}_{esc}} \approx 0$; in this case, we can observe a maximum peak at $\omega=0$. which also corresponds to a large variation.
Finally, if $\tau^{'}_{esc} \approx \tau^{''}_{esc}$, that is, $\omega = \frac{1}{2}$, which is the condition for small variation in the first escape time between the trajectories.

Finally, to resolve the microscopic displacement statistics and identify transient active trapping within the corrugated channels, we compute the self-part of the van-Hove correlation function (VHCF) \cite{simma2025competing, Chen_Nanrong_2022, Ghosh_2016, kim2022, yin2021_Frontiers_of_Physics}. Because the corrugated geometries induce anisotropic transport, we evaluate the VHCF independently along the longitudinal ($x$) and transverse ($y$) directions. The longitudinal VHCF is mathematically defined as:

\begin{equation} \label{vanHove_x_equation}
 G_s (\Delta x; \tau) \equiv \langle \delta (\Delta x - (x(t+\tau) - x(t)))\rangle,
 \end{equation}

where $\Delta x$ is the spatial displacement over a specific lag time $\tau$. The transverse function, $G_s (\Delta y; \tau) \equiv \langle \delta (\Delta y - (y(t+\tau) - y(t)))\rangle$, is defined analogously.

%==========================================================================

\section{Results}
\subsection{Radial probability distribution function (RPDF), $P(r)$  }
%=========================================================================
\begin{figure*}
    \includegraphics[width=\textwidth]{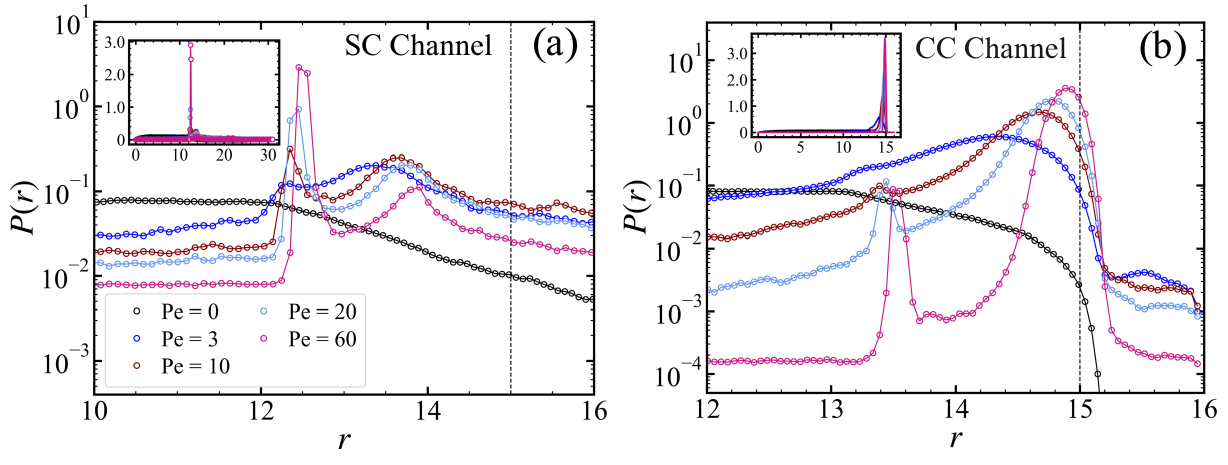}
  \caption{Semi-logarithmic radial probability distribution function, $P(r)$, of the active rod prior to its escape from (a) a SC channel and (b) a CC channel, for different activity strengths ($Pe$). The inset displays the corresponding full probability distribution on a linear scale. The vertical dashed line at ($r=15\sigma$) indicates the maximum distance in the transverse direction of the boundary ($y_{m}$) of the SC channel, providing a geometric reference for boundary accumulation. Here all the distances measured in units of $\sigma$, the fundamental unit of length.}
  \label{RDF_Plot}
\end{figure*}
%=========================================================================

To investigate the rod's interaction with the channel topology, we first compute the radial probability distribution function (RPDF), $P(r)$. For comparison, we initially consider $P(r)$ within the sinusoidal channel; see Fig. \ref{RDF_Plot}(a). The rod is released vertically and given sufficient equilibration time to minimize initial bias. We observe that the RPDF of the passive rod ($Pe = 0$) exhibits a nearly uniform distribution across all radial distances within the sinusoidal blob ($r < 30\sigma$); see Fig. \ref{Model}(a,c). The random thermal fluctuations of passive rod dominated the translational movement; as a result, the rod uniformly traces the inside region of the sinusoidal blob; see Fig. \ref{Active_Rod_Trj}(b) for random trajectories. Due to the smooth curvature of the sinusoidal channel, the probability decays gradually at the boundary. Notably, the escape of the passive rod through the channel neck is an exceedingly rare event, and none were observed within our simulation time frame. 

As activity is introduced (\(Pe>0\)), a distinct bimodality emerges in the distribution profiles, and directed motion facilitates escape events; see Fig. \ref{Active_Rod_Trj}(c). This bimodality in \(P(r)\) arises from the rod colliding with and subsequently aligning along the channel wall. Given that the half-vertical length of the channel is \(15\sigma\); see Fig. \(\ref{Model}(a)\), the first peak occurs at a radial distance of roughly \(12.5\sigma\). This peak is associated with rod–wall collisions near the channel maxima. Because the rod has a finite length, its center of mass remains offset from the wall when the rod tip collides with the boundary, resulting in a peak approximately \(2.5\sigma\) from the wall. This interpretation reflects the combined effects of rod geometry, orientational dynamics, and confinement. The collision-induced peak becomes more prominent at higher activities (\(Pe>10\)), when persistent propulsion enhances interactions with the channel boundaries. Following a collision, the rod reorients in response to the local channel geometry and subsequently moves along the boundary toward the escape region. During this process, the rod frequently adopts a near-vertical orientation, giving rise to the enhanced peak around \(\theta\approx\pi/2\) (measured relative to the \(x\)-axis) in the SC channel, as shown by the angular probability distribution in Fig. \ref{PDF_Angles}. At lower propulsion strengths, the rod requires less time to align with the wall, resulting in a diminished first peak. Once properly oriented, the rod slides along the channel's curvature toward the neck, generating a second, more pronounced peak at $r \approx 14\sigma$. The uniform spatial distribution seen in the passive case is thus replaced by strong boundary accumulation. To further support this, the MSD for various activities is provided in the supplementary material Fig. \ref{MSD_SI}(a).

Having established the dynamics in the smoothly curved SC channel, we now turn to the highly curved CC channel, see Fig. \ref{Model}(b). Sample trajectories are shown in Fig. \ref{Model}(d), and the corresponding RPDF is plotted in Fig. \ref{RDF_Plot}(b), with the inset depicting the full radial distance prior to escape.

As depicted in Fig. \ref{Model}(b), given the circular blob's radius of $16\sigma$, the passive rod uniformly explores the interior space ($r < 16\sigma$) prior to escape. Because the dynamics are driven purely by thermal fluctuations, this exploration yields a uniform $P(r)$ profile, followed by a sharp decay at the boundary due to wall repulsion; see the black curve in Fig. \ref{RDF_Plot}(b). Escape events remain extremely rare.

For the self-driven rod ($Pe > 0$) in the CC channel, we observe a similar bimodal profile, though the relative intensities of the peaks differ. The first collision peak occurs at $13.5\sigma$. However, because the higher curvature of the CC channel allows the rod to align with the wall much faster than in the sinusoidal channel, this first collision peak is significantly less intense. Following this rapid initial hit, the rod orients and slides along the boundary, pushing its center of mass closer to the wall. This results in a dominant second peak near $r \approx 15\sigma$, which is almost close to the boundary.

The higher intensity of the second peak indicates that circular geometry facilitates faster alignment of the rod along the boundary. Once oriented, the rod traces the boundary curvature, remaining trapped at the wall for an extended period before achieving the proper orientation to escape. Furthermore, the MSD of the self-driven rod inside the CC channel is provided in the supplementary material Fig. \ref{MSD_SI}(b).

\subsection{MFET, CV of FET}

%=========================================================================
\begin{figure}
    \includegraphics[width=8cm]{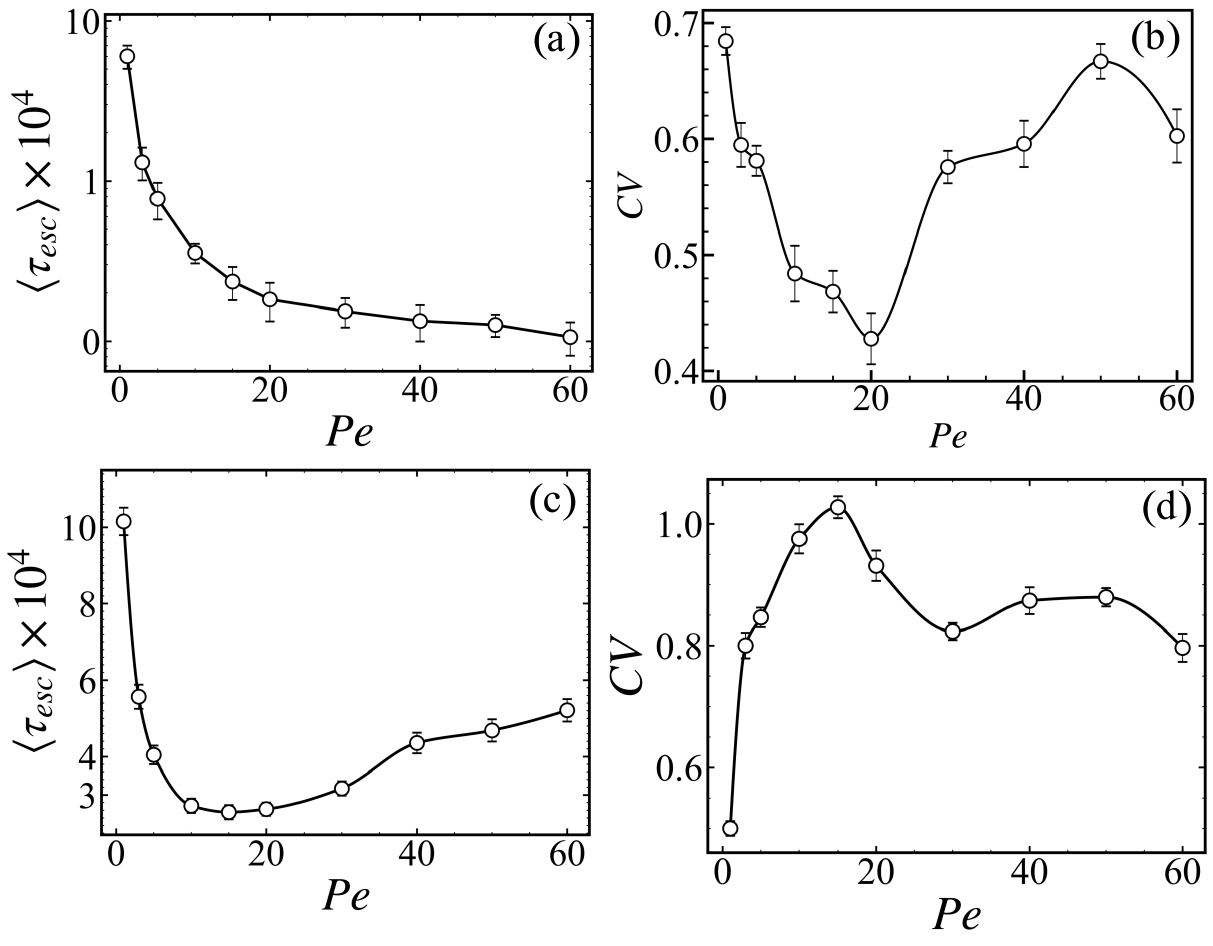}
  \caption{Escape kinetics and associated statistical fluctuations of the active rod. (a, c) Mean first escape time (MFET) versus $Pe$ in (a) SC channel, and (c) CC channel. (b, d) The corresponding coefficient of variation (CV) of first escape times as a function of P\'{e}clet number.}
  \label{FET_CV_Plot}
\end{figure}
%=========================================================================

The escape behavior of the rod can be assessed using the mean first escape time (MFET) and the coefficient of variation (CV) of the escape times, defined in Eq. \ref{MFET_equation} and Eq. \ref{CV_equation}, respectively. Fig. \ref{FET_CV_Plot}(a) presents the MFET as a function of the propulsion strength ($Pe$) for the self-driven rod in the sinusoidal corrugated (SC) channel. Being consistent with previous observations in RPDF, the passive rod ($Pe = 0$) performs a random trajectory and struggles to find the opening, taking a practically infinite time to escape. However, for $Pe > 0$, inherent activity drives directed motion, and the MFET monotonically decreases as $Pe$ increases. The smooth geometry of the SC channel effectively guides this directed motion toward the neck, which triggers faster escape events. Consequently, for SC channel, highly self-driven rods ($Pe > 20$) escape significantly faster than those with low activity. 

In contrast, the highly curved CC channel exhibits a non-monotonic MFET trend as a function of $Pe$, as shown in Fig. \ref{FET_CV_Plot}(c). At lower propulsion strengths ($Pe < 20$), the rod spends most of its time in the bulk region of the circular blob; see Fig. \ref{Model}(d). While it may orient along the direction of channel opening, its directed speed is insufficient to quickly pass through the narrow channel neck ($w = 1.5\sigma$). At an optimal intermediate propulsion ($Pe \approx 20$) yields the minimum MFET, as the channel topology efficiently guides the moderately active rod into the proper orientation for escape through neck. As propulsion is increased further ($Pe > 20$), we notice a significant increase in the MFET. At these high activities, the rod strongly aligns and sticks to the wall of the circular blob, unable to reorient toward the exit. Unlike the SC channel, the converging geometry near the CC channel's neck enhances steric confinement, obstructing the passage of fast-moving rods and resulting in prolonged confinement times.

To capture the statistical variations in available set of escape events, we evaluate the Coefficient of variation (CV) of the first escape times. Fig. \ref{FET_CV_Plot}(b) and (d) show the CV versus $Pe$ for the SC and CC channels, respectively. In the SC channel, the CV is large at lower propulsions ($Pe < 5$) because wall collisions generate a wide variance in escape times. At intermediate activities ($Pe \approx 10, 20$), the escape events occur at more consistent intervals, yielding a lower CV. At higher $Pe$, outliers emerge again, resulted in an increment in the CV. Conversely, in the CC channel, fluctuations are minimized at lower propulsions ($Pe < 10$), where outliers are extremely rare. The CV is maximum at $Pe \approx 10$, indicating a maximum in escape time fluctuations, before gradually decreasing at higher $Pe$. The CV and MFET characterize different aspects of escape dynamics, with the CV reflecting trajectory-to-trajectory fluctuations that depend on channel geometry. In the SC channel, fluctuations increase at higher activity, whereas in the CC channel, competing escape pathways and prolonged wall interactions enhance fluctuations at intermediate activity, highlighting the role of channel morphology and particle–wall interactions\cite{Souzy_PRR_2022, caprini_PRR_2020}. Although the CV provides a useful quantitative measure of these fluctuations, a more comprehensive understanding requires examining the full probability distributions of the escape times, which we discuss in the following section. 

%===============================================
\subsection{PDF of FET, Uniformity index}

%=========================================================================
\begin{figure*}
    \includegraphics[width=\textwidth]{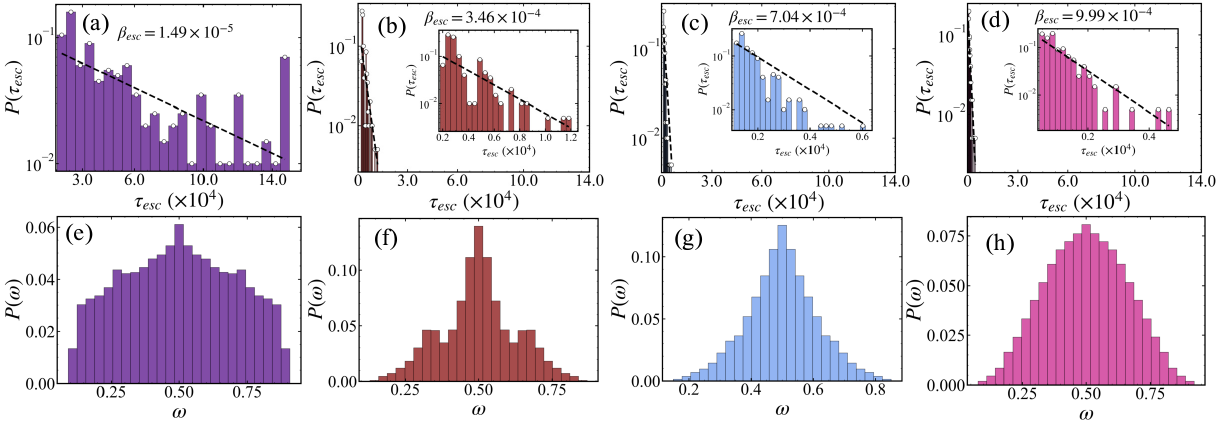}
  \caption{Probability distribution function of the first escape time of the active rod in the SC channel. The top row displays the $P(\tau_{esc})$, while the bottom row presents the corresponding uniformity index, $P(\omega)$. Colors from left to right correspond to increasing active propulsion strengths ($Pe = 1, 10, 20, 60$). Insets highlight the explicit decay of the distributions. Black dashed lines denote exponential fits; from the slopes, we calculated the decay constant $\beta_{esc}$).}
  \label{PDF_FET_UI_Sin_CC}
\end{figure*}
%=========================================================================

Based on our analysis of the MFET, CV, we now investigate the first escape times (FETs) and their fluctuations in a probabilistic context. Fig. \ref{PDF_FET_UI_Sin_CC} presents the probability density function (PDF) of the first escape time, $P(\tau_{esc})$, and the corresponding PDF of the uniformity index, $P(\omega)$, for various activity levels in the SC-channel. The black dashed lines represent the exponential fit $\log(P(\tau_{esc})) = a + \beta_{esc} \tau_{esc}$, where $\beta_{esc}$ is the decay constant defining the rate of the escape process.

For a low activity of $Pe = 1$ (Fig. \ref{PDF_FET_UI_Sin_CC}(a, e), $P(\tau_{esc})$ decays slowly with a small decay constant, indicating that the rod takes a longer time to escape. The corresponding $P(\omega)$ exhibits a broad, nearly equal probability across all $\omega$ values, reflecting significant trajectory-to-trajectory fluctuations in the escape times. As propulsion increases to $Pe = 3$ (supplementary see Fig. \ref{PDF_FET_UI_Sin_CC_SI}(a), inset for a clearer view of the decay), the distribution shifts toward lower $\tau_{esc}$ values, exhibiting a sharp, rapid decay. This confirms that increased propulsion helps the rod escape faster. Simultaneously, $P(\omega)$ develops a pronounced peak at $\omega = 1/2$, clearly indicating that escape times become highly consistent (fewer fluctuations) across different trajectories. This rapid decay persists for intermediate propulsion strengths ($Pe = 10, 20$; see Fig. \ref{PDF_FET_UI_Sin_CC}(b,c,f,g) histograms), where effective orientation leads to early escape events and strongly peaked $P(\omega)$ distributions.

As propulsion increases further ($Pe \approx 60$); see Fig. \ref{PDF_FET_UI_Sin_CC}(d, h), $P(\tau_{esc})$ still exhibits a sharp decay towards low escape times, but $P(\omega)$ slightly diverges from $\omega = 1/2$. At this high activity, while the rod eventually escapes rapidly, the increased number of wall collisions introduces higher trajectory-to-trajectory variance, consistent with the CV behavior discussed previously, Fig. \ref{FET_CV_Plot}(b); see supplementary Fig. \ref{PDF_FET_UI_Sin_CC_SI} for intermediate values not shown in the main text.
%=========================================================================
\begin{figure*}
    \includegraphics[width=\textwidth]{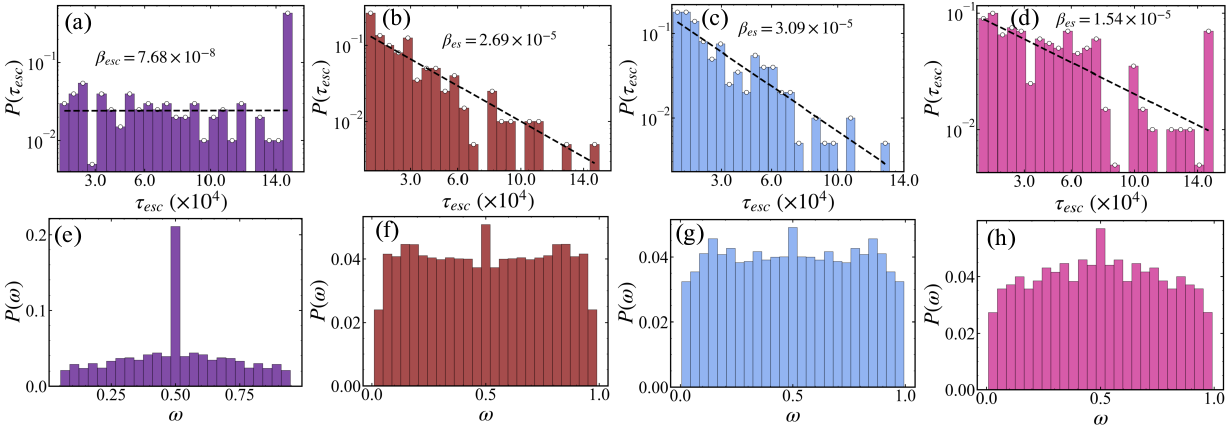}
  \caption{Probability distribution function of the first escape time in the circular corrugated (CC) channel. The top row displays $P(\tau_{esc})$, while the bottom row presents the corresponding uniformity index, $P(\omega)$. The color profile corresponds to increasing propulsion strengths ($Pe = 1, 10, 20, 60$), adhering to the color scheme established in Fig.~\ref{PDF_FET_UI_Sin_CC}. Black dashed lines denote exponential fits to the data, the slopes of which yield the decay constant ($\beta_{esc}$).}
  \label{PDF_FET_UI_dia_32}
\end{figure*}
%=========================================================================

We now turn our attention to the CC channel depicted in Fig. \ref{Model}(d) to highlight the differences in escape dynamics of rod relative to the SC-channel. Fig. \ref{PDF_FET_UI_dia_32} shows $P(\tau_{esc})$ and the corresponding $P(\omega)$ for the CC channel. At low propulsion ($Pe = 1$, indigo), the strong confinement of the circular blob severely delays escape. We observe a peak intensity in $P(\tau_{esc})$ at approximately $\tau_{esc} = 15 \times 10^4$. Because almost all trajectories require this extended time to escape, $P(\omega)$ is strongly peaked at $\omega = 1/2$, indicating minimal trajectory-to-trajectory fluctuations, which is consistent with the CV data in Fig. \ref{FET_CV_Plot}(d).

As activity increases from $Pe = 3$ to $Pe = 20$, the decay constant $\beta_{esc}$ monotonically increases, reaching a maximum at the optimal activity of $Pe \approx 20$. In this moderate activity regime, the rod has sufficient speed and orientation to efficiently navigate the pore. Interestingly, for these intermediate activities, $P(\omega)$ spans the full range from $\omega = 0$ to $1$ without a pronounced peak. This broad distribution provides strong evidence that the escape times are nearly follow uniform distribution, signifying high trajectory-to-trajectory fluctuations.

Beyond this optimal propulsion, a change in the underlying escape mechanism occurs. For $Pe \approx 60$; Fig. \ref{PDF_FET_UI_dia_32}(e, j), we observe a significant decrease in the decay constant. Despite its high propulsion speed, the rod struggles to achieve a suitable escape orientation, as it persistently traces the circular boundary. For this high activity, $P(\omega)$ once again converges around $\omega = 1/2$, unlike the moderately active cases. This suggests that excessively strong propulsion in a highly curved geometry suppresses trajectory-to-trajectory fluctuations, as all rods experience similar, deterministic trapping along the boundary before eventually escaping; see Fig. \ref{PDF_FET_UI_Sin_CC_SI} for {intermediate values not present in the main text.}

\begin{figure}
\includegraphics[width=8cm]{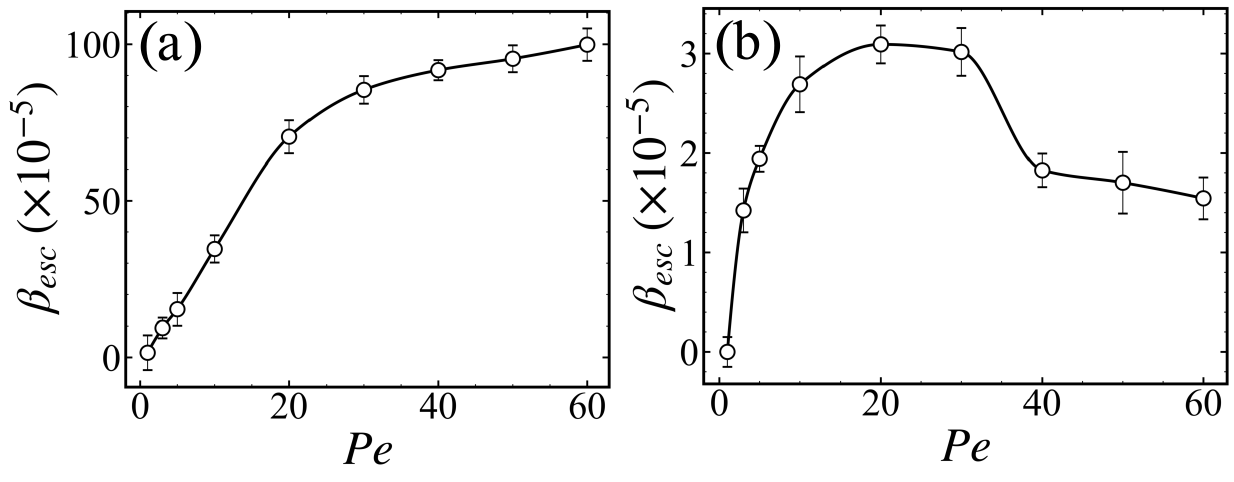}
\caption{Variation of decay constant with active propulsion strength $Pe$. The decay constant ($\beta_{esc}$) is shown as a function of the P'{e}clet number ($Pe$) for the active rod in (a) the sinusoidal corrugated (SC) channel and (b) the circular corrugated (CC) channel. Solid lines connecting the data points are for guidance to the eye.}
\label{DecayConstant_Plot}
\end{figure}

To systematically quantify the escape kinetics, Fig. \ref{DecayConstant_Plot} presents the decay constant ($\beta_{esc}$) as a function of the propulsion strength ($Pe$). For the SC channel Fig. \ref{DecayConstant_Plot}(a), $\beta_{esc}$ increases monotonically with activity up to $Pe \approx 20$, beyond which a noticeable saturation occurs. In this geometry, the smooth channel curvature efficiently guides the aligned self-driven rod toward the exit, leading to monotonically faster escape kinetics that plateau at high propulsion speeds. Conversely, the rod dynamics within the CC channel, Fig. \ref{DecayConstant_Plot}(b), exhibit a distinct non-monotonic dependence of $\beta_{esc}$ on $Pe$. The decay constant reaches a maximum at an optimal propulsion strength ($Pe \approx 20$), indicating the most favorable conditions for rapid escape. At higher activities ($Pe > 20$), despite the increased self-driven speed, the self-driven rod becomes persistently trapped along the highly curved circular boundary. This boundary accumulation prolongs the confinement time, resulting in the observed reduction of the decay constant.
%================================================================================

%\subsection{Decay Constant Vs Propulsion}

%=========================================================================

%=========================================================================

\subsection{van-Hove correlation function}

Typically, the van-Hove correlation function (VHCF) of an active rod in free space exhibits a crossover from Gaussian behavior at short and long time scales to non-Gaussian behavior in the intermediate super-diffusive regime; see supplementary Fig. \ref{vanHove_SI} and Fig. \ref{MSD_RMSD_freespace_SI}(a). The Gaussian profile, characteristic of normal Fickian diffusion \cite{Smoluchowski1906, Chandrasekhar1943, wang2012brownian}, can be represented by the analytical form:

\begin{equation} \label{gaussian_equation} 
G_s(\Delta x; \tau) \frac{1}{\sqrt{2\pi \langle \Delta x^2 \rangle}}\exp\left(-\frac{\Delta x^2}{2 \langle \Delta x^2 \rangle}\right),
\end{equation}

where $\Delta x$ is the displacement over a specific lag time $\tau$. The emergence of intermediate super-diffusion, which yields non-Gaussian profiles, is well documented in active systems \cite{Weeks2000, Chaudhuri2007}. A representative Gaussian fit for $\tau = 50$ is shown in Fig. \ref{vanHove_gaussian_fit} supplementary.

\begin{figure}
    \includegraphics[width=8cm]{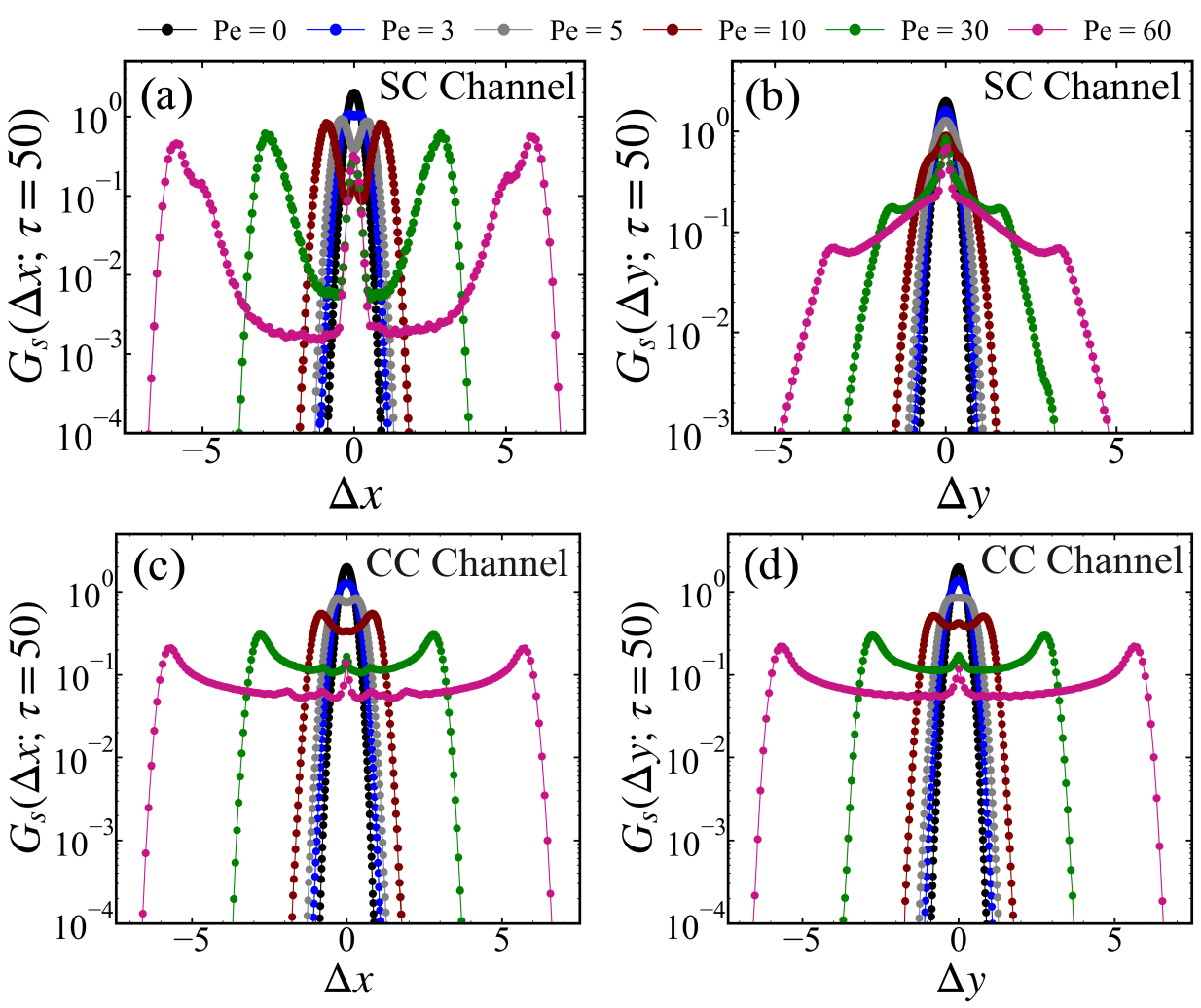}
  \caption{The self part of the van-Hove correlation function (VHCF) of the self-driven rod in the SC channel (top row) and the CC channel (bottom row) at a fixed lag time $\tau = 50$ for the different values of the P\'{e}clet number($Pe$). (a, c) are for the x-direction, and (b, d) represent the y-direction. Solid lines connect the data points as guides to the eye. }
  \label{vanHove_Plot}
\end{figure}

Fig. \ref{vanHove_Plot} presents the self-part of the VHCF for the sinusoidal corrugated (SC) and circular corrugated (CC) channels in both the $x$- and $y$-directions at a fixed lag time ($\tau = 50$). For the passive rod ($Pe = 0$), the VHCF strictly follows the Gaussian distribution (Eq. \ref{gaussian_equation}) in both directions and across both channel geometries, reflecting purely thermal, random diffusion; see Fig. \ref{MSD_SI}(a). However, introducing activity ($Pe > 0$) induces significant non-Gaussianity due to interactions between the active rod and the channel topology.

%=========================================================================
%=========================================================================

In the highly anisotropic SC channel Fig. \ref{vanHove_Plot}(a, b), low propulsion ($Pe < 10$) yields a central dip in the VHCF along the $x$-direction. The weakly self-driven rod smoothly aligns with the wall, restricting its longitudinal displacement. At higher propulsion ($Pe \geq 10$), strong collisions with the rigid boundary generate a prominent central peak. Because highly self-propel rods traverse larger distances within the lag time, they subsequently slide along the sinusoidal wall toward the pore, generating broad, non-Gaussian heavy tails at larger displacements Fig. \ref{vanHove_Plot}(a). Conversely, along the confined $y$-direction Fig. \ref{vanHove_Plot}(b), transverse displacement is sterically limited. Consequently, the central peak intensity is nearly independent of $Pe$, and wall-sliding manifests as a smooth, flattened decay rather than extended tails\cite{yin2021_Frontiers_of_Physics}. {The van-Hove profiles corresponding to the lag time $\tau \approx 200$ given to demonstrate the non-Gaussian behavior; see Fig. \ref{vanHove_lt_200_SCC} in supplementary.}

Unlike the SC channel, the CC channel enforces symmetric confinement within its circular blob Fig. \ref{vanHove_Plot}(c, d). Thus, the self-driven rod exhibits similar non-Gaussian VHCF profiles in both the $x$- and $y$-directions. For $Pe > 10$, a sharp central peak emerges at small displacements due to boundary collisions. Beyond this peak, the VHCF exhibits an extended plateau with nearly equal probabilities for intermediate displacements. This plateau reflects active boundary trapping: upon hitting the wall, the rod persistently slides along the isotropic circular curvature for extended periods before escaping. The spatial extent of this equal-probability region scales monotonically with activity, as faster rods explore larger arc lengths along the boundary during the fixed lag time; see Fig. \ref{MSD_SI}(b) for MSD.
The non-Gaussian behavior still prevails for higher lag times as the rod influenced by confined geometry; see Fig \ref{vanHove_lt_200_CCC} for van-Hove profiles of the rod at lag time $\tau \approx 200$ in CC channel.

Overall, the channel curvature fundamentally governs the non-equilibrium displacement statistics of the self-driven rod. The smoothly varying, anisotropic SC channel produces broad tails in the longitudinal direction and sterically truncated profiles transversely. In contrast, the highly curved, symmetric CC channel promotes active trapping, yielding extended plateaus of uniform probability in both spatial directions.

%======================================================================================
\section{Discussion and Conclusion}
\label{sec-conclusion}

\subsection{Summary of the main results}

In summary, we investigated the escape dynamics of self-driven rods {confined within narrow} corrugated channels. To elucidate the topological effects, we compared two distinct {confinement} geometries: {a smoothly varying sinusoidal channel (SC) and a circular corrugated (CC) channel of fixed uniform curvature.} 

{Previous studies of active particles escaping through entropic barriers have reported that the mean first-escape time (MFET) can decrease with increasing propulsion strength \cite{Biswas2020, Biswas_Kudrolli_2023, Upadhyaya2024}.} {However, the dependence of escape dynamics on activity is not universal. In particular, studies that characterize activity through the persistence time, typically defined as the inverse of the rotational diffusion coefficient, have reported non-monotonic escape-time behavior arising from the competition between persistent motion and particle-boundary interactions \cite{Caprini_JCP_2019, angelani_EPJE_2024}.} In this context, our study extends this framework to anisotropic active particles inside corrugated geometries with different topological constraints. We demonstrate that the escape behavior is not only determined by the propulsion strength but also {critically by the local channel morphology} which plays a crucial role, and most importantly, we have shown that, depending on the geometry, a higher propulsion strength may hinder the escape dynamics.

In both geometries, the radial probability density function, $P(r)$, exhibits a distinct bimodal distribution due to the persistent collision and subsequent alignment of the rods along the {rigid boundaries}. Despite this {shared bimodality}, the underlying orientational and boundary-association kinetics differ significantly. The smoothly varying curvature of the SC channel induces a slower orientational alignment but longer association with the wall, whereas the sharper curvature of the CC channel forces rapid boundary alignment but faster dissolution from the wall. Consequently, the MFET manifests distinct topological dependencies. In the SC channel, the MFET decreases monotonically with $Pe$, where as the CC channel induces a strictly non-monotonic dependence, revealing an optimal propulsion strength for an efficient escape mechanism.

%===========================
The statistical variation in FETs further supports the topological dependency. In SC channel, the probability density function of the FETs, $P(\tau_{esc})$, exhibits an exponential decay with respect to $\tau_{esc}$, where the corresponding decay constant increases monotonically with $Pe$. While $P(\tau_{esc})$ in the CC channel follows a similar exponential decay profile, and its decay constant exhibits a strictly non-monotonic dependence on propulsion strength. This transition from predictable boundary sliding to intermittent active trapping is reflected in the distribution of the uniformity index ($\omega$). In the SC channel, the probability density of $\omega$ sharply peaks at $\omega \approx 1/2$ for intermediate propulsion strengths, reflecting minimal trajectory-to-trajectory variance in escape times. Conversely, the distribution of the uniformity index for the CC channel diverges toward the extremes, reflecting highly stochastic, trajectory dependent escape kinetics. Finally, the microscopic displacement statistics captured by the van-Hove correlation function (VHCF) reflect the underlying channel symmetries: the anisotropic geometry of the SC channel generates distinct non-Gaussian profiles in the longitudinal ($x$) and transverse ($y$) directions, whereas the isotropic CC cavity enforces symmetric VHCF profiles featuring extended plateaus indicative of persistent boundary trapping along both axes.

%=================

\subsection{Application and perspective}

Our findings show that precise control over the geometry and topology of the corrugated region can dictate the transport and escape dynamics of active nanorods. The orientational trapping and alignment-assisted sliding elucidated here closely parallel the boundary tracing navigation strategies used by motile biological microorganisms, such as bacteria and spermatozoa, as they travel through mucosal networks or reproductive tracts \cite{ELGETI_BiophysicalJournal_2010, Celli_PNAS_2009}. Furthermore, the robust coupling between local curvature and transport efficiency establishes quantitative principles that have direct implications for synthetic systems. Specifically, this understanding can help design artificial micro-environments engineered for selective particle separation and sorting \cite{McGrath_DLD_2014, Bechinger2016}. Our current study potentially offers a physical mechanism relevant to controlled transport in microfluidic ratchets or nonporous matrices \cite{Reguera2006, Agosta_Mikhail_JCP_2024} and targeted cargo delivery by catalytic nanomotors \cite{Ma_JACS_2015, Wang_ChemRev_2015}. {Future studies can be designed the generalized framework of interacting active rods to investigate how particle density, steric hindrance, and asymmetric channel geometry jointly influence collective dynamics and large-scale transport. Increasing density may promote orientational correlations and wall accumulation while obstructing narrow openings, potentially leading to clogging and modified escape dynamics. Systematic exploration of these competing effects as functions of density, aspect ratio, activity, and channel geometry would provide further insight into collective transport in confined active matter.}

\section{Acknowledgements}
SKG would like to thank MATRICS, ANRF (Project No. MTR/2022/000833) for financial support. The authors acknowledge the computational resources  PARAM Rudra and PARAM Shakti Cluster from the National Supercomputing Mission (NSM), Government of India. We want to acknowledge NIT Warangal for the fellowship to BR. SKG would like to thank Prof. Supravat Dey for useful discussions.   

%\section{TO DO}
%\begin{enumerate}
%    \item \red{\st{Put Error bars in Fig.3, Fig.6.} is implemented}
%    \item \red{\st{Check Supplementary Materials} checked it}
%\end{enumerate}

\bibliographystyle{unsrt}
\section*{References}
\bibliography{Rod_in_CC.bib}

\clearpage

\appendix
\counterwithin{figure}{section}

\onecolumngrid

\section{ACRONYMS}

\begin{table}[htb]
\centering
\label{tab:abbreviations}

\large
\renewcommand{\arraystretch}{1.5}
\setlength{\tabcolsep}{20pt}

\begin{tabular}{|l|l|}
\hline
\textbf{Acronym} & \textbf{Full form} \\ \hline
CC & Circular Corrugated \\
SC & Sinusoidal Corrugated \\
RPDF & Radial Probability Distribution Function \\
MFET & Mean First Escape Time \\
FET & First Escape Time \\
CV & Coefficient of Variation \\
VHCF & van-Hove Correlation Function \\
\hline
\end{tabular}
\end{table}

%=====================================================

\section{SIMULATION PARAMETERS}

\begin{table}[htb]
\centering
\label{tab:simulation_parameters}

\large
\renewcommand{\arraystretch}{1.5}
\setlength{\tabcolsep}{20pt}

\begin{tabular}{|l|l|l|}
\hline
\textbf{Parameter} & \textbf{Symbol} & \textbf{Value} \\ \hline
Particle diameter & $\sigma$ & 1.0 \\
Particle mass & $m$ & 1.0 \\
Thermal energy & $k_{\mathrm{B}}T$ & 1.0 \\
Lennard--Jones energy & $\varepsilon$ & $1.0\,k_{\mathrm{B}}T$ \\
Time step & $\Delta t$ & 0.0005 \\
Rod length & $L$ & $5\sigma$ \\
Simulation time & $T$ & $3 \times 10^8$ \\
\hline
\end{tabular}
\end{table}
%=======================================================
\clearpage

%=========================================================================
\section{SUPPLEMENTARY FIGURES}

\begin{figure*}[htb]
    \includegraphics[width=\textwidth]{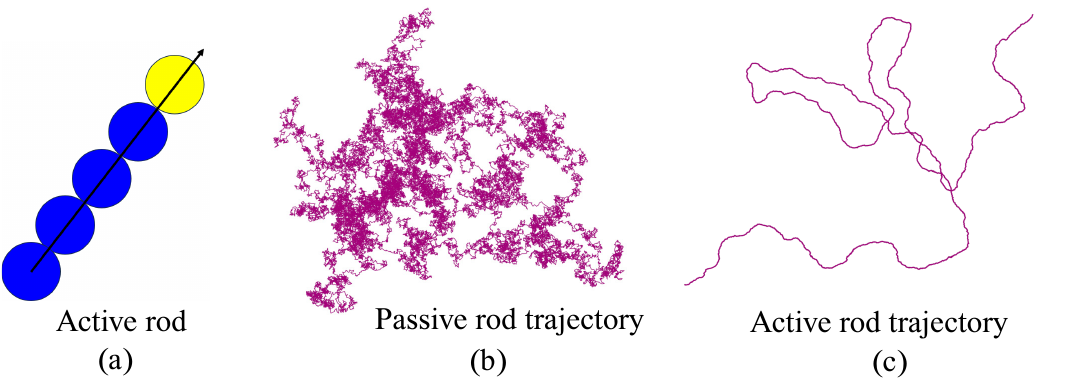}
  \caption{Schematic of the active rod model and the corresponding trajectories in free space. (a) Illustration of the active nanorod, where the yellow bead denotes the head monomer and the black arrow indicates the axis of self-propulsion. (b, c) Representative center of mass trajectories for (b) a purely passive rod ($Pe = 0$) and (c) an active rod ($Pe = 5$) in free space. The passive trajectory exhibits standard Brownian motion, whereas the active trajectory demonstrates persistent directed motion coupled with rotational diffusion.}
  \label{Active_Rod_Trj}
\end{figure*}
%=========================================================================

%=========================================================================
\begin{figure*}
    \includegraphics[width=\textwidth]{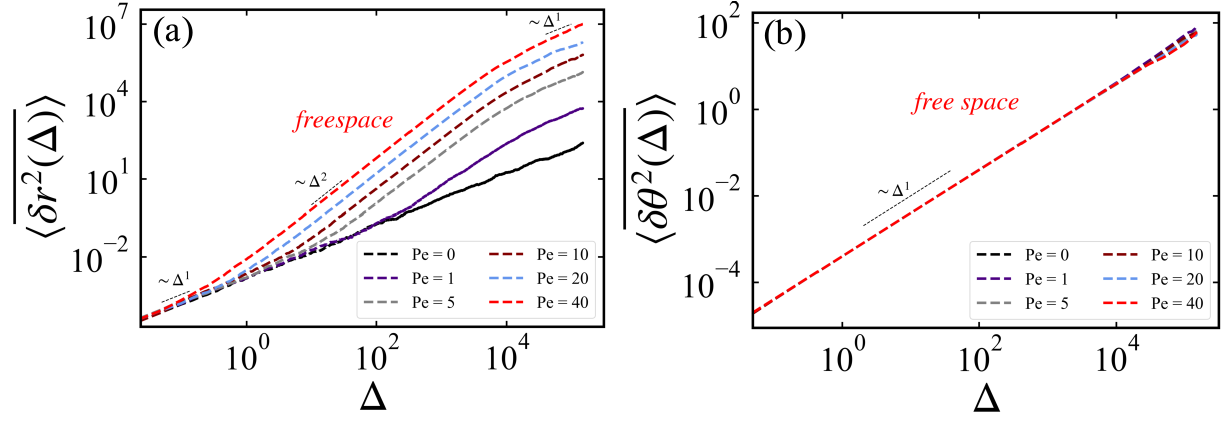}
  \caption{Translational and rotational dynamics of active rod. (a) Translational mean squared displacement ($MSD$) and (b) rotational mean squared displacement of the active rod in free space. Black dashed lines indicate theoretical power-law scalings.}
  \label{MSD_RMSD_freespace_SI}
\end{figure*}
%=========================================================================

%=========================================================================
\begin{figure*}
    \includegraphics[width=\textwidth]{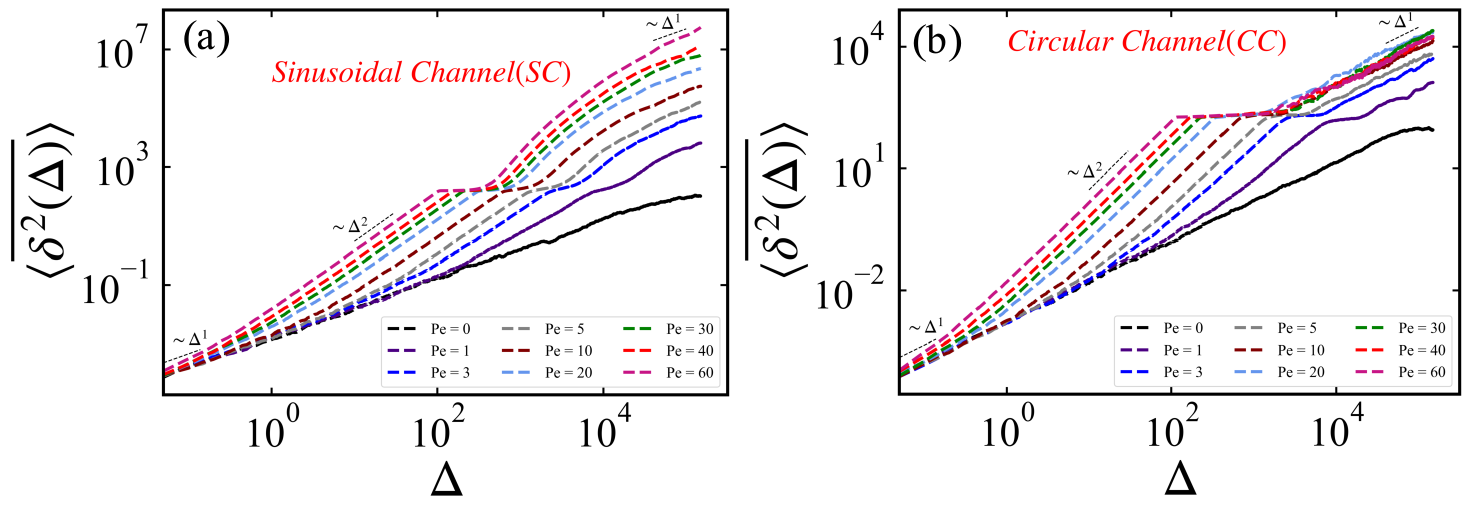}
  \caption{Time-averaged mean squared displacement (MSD) of the confined active rod in different channel geometries: (a) sinusoidal corrugated (SC) channel and (b) circular corrugated (CC) channel. Dashed lines denote theoretical power-law scalings.}
  \label{MSD_SI}
\end{figure*}
%=========================================================================

%=========================================================================
\begin{figure*}
    \includegraphics[width=15cm]{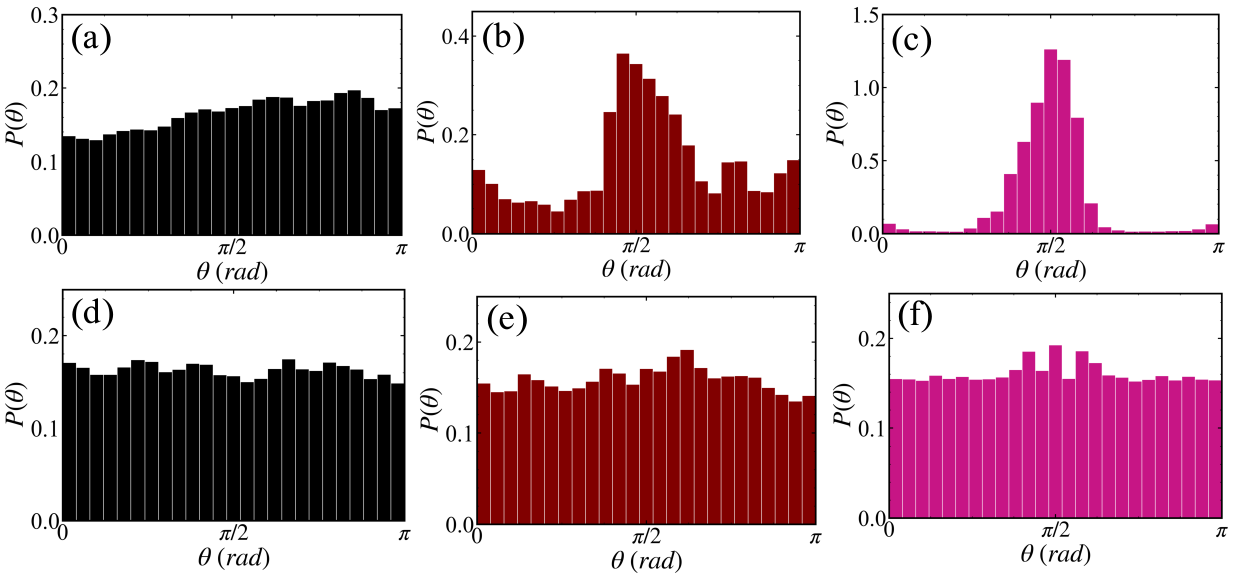}
  \caption{Probability distribution function of the active rod orientation angle, $P(\theta)$, for different Péclet numbers (Pe). The top row (a–c) displays the angular distributions for rods confined in a SC channel at $Pe = 0, 10$, and $60$, respectively. The bottom row (d–f) shows the corresponding angular distributions for rods in a CC channel at $Pe = 0$, $10$, and $60$, respectively. As activity increases in the straight channel, the rods transition from a uniform random orientation to a strongly aligned state parallel to the walls, indicated by the sharp peaks at $\theta = 0$ and $\theta = \pi$. Conversely, high activity in the corrugated channel promotes distinct orientation preferences, notably perpendicular to the main axis near $\theta = \pi/2$.}
  \label{PDF_Angles}
\end{figure*}
%=========================================================================

%=========================================================================
\begin{figure*}
    \includegraphics[width=\textwidth]{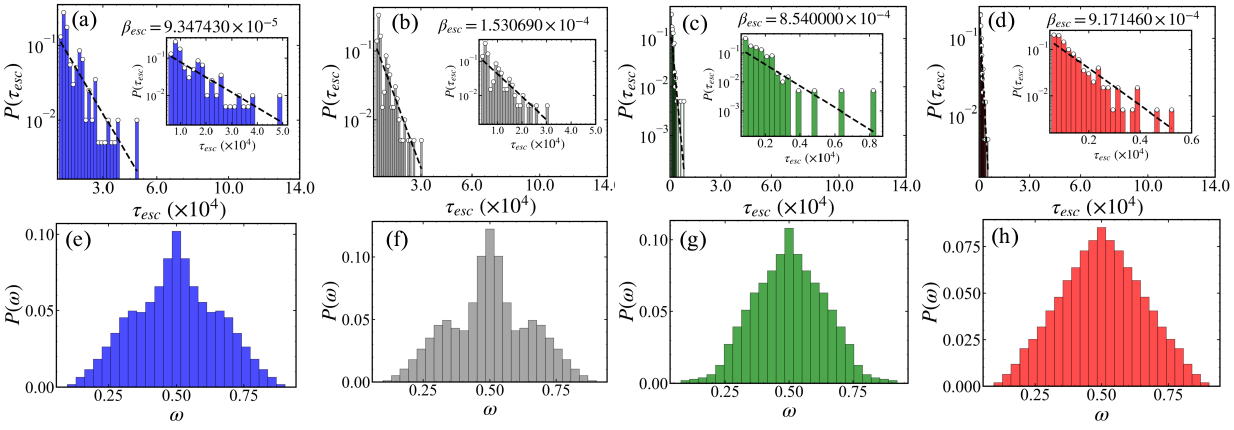}
  \caption{Probability density functions (PDF) of the first escape times, $P(\tau_{esc})$, and the corresponding uniformity index, $P(\omega)$, for the active rod in the sinusoidal corrugated (SC) channel. Results are shown for four distinct P\'{e}clet numbers: $Pe = 3$ (blue), $5$ (gray), $30$ (green), and $40$ (red). Dashed lines indicate exponential fits to the long-time tails of the escape time distributions, the slopes of which yield the macroscopic decay constant ($\beta_{esc}$).}
  \label{PDF_FET_UI_Sin_CC_SI}
\end{figure*}
%=========================================================================

%=========================================================================
\begin{figure*}
    \includegraphics[width=\textwidth]{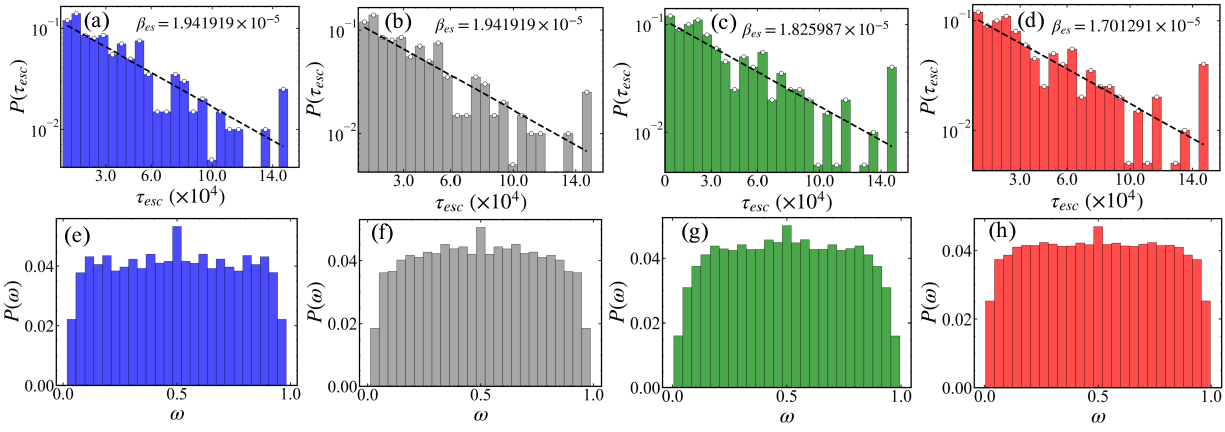}
  \caption{Probability density functions (PDF) of the first escape times, $P(\tau_{esc})$, and the corresponding uniformity index, $P(\omega)$, for the active rod in the circular corrugated (CC) channel. Results are shown for four distinct P\'{e}clet numbers: $Pe = 3$ (blue), $5$ (gray), $30$ (green), and $40$ (red). Dashed lines indicate exponential fits to the long-time tails of the escape time distributions, the slopes of which yield the macroscopic decay constant ($\beta_{esc}$).}
  \label{PDF_FET_UI_dia_32_SI}
\end{figure*}
%=========================================================================

%=========================================================================
\begin{figure*}
    \includegraphics[width=12
    cm]{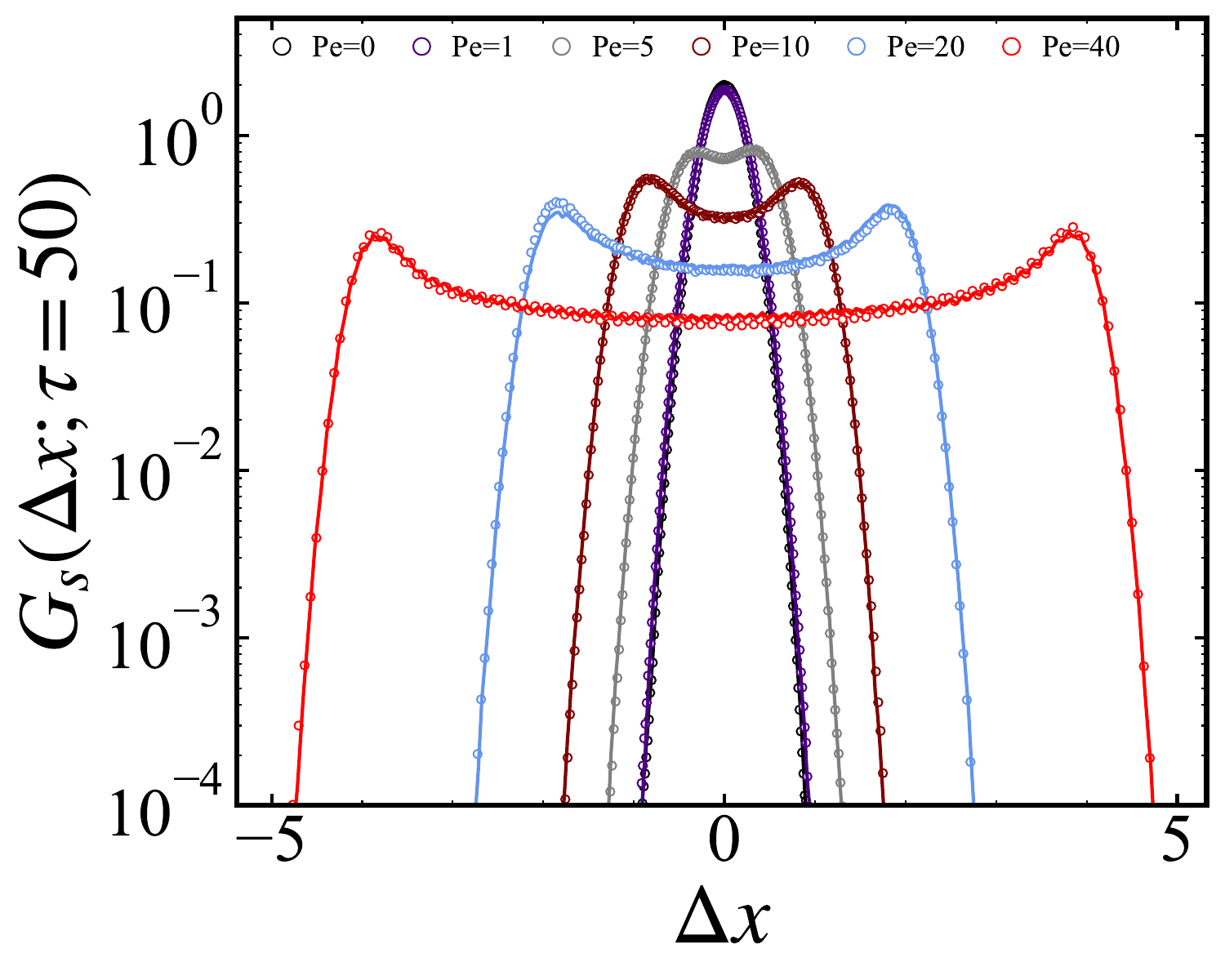}
  \caption{Self-part of the van-Hove correlation function for the unconfined active rod. The displacement distribution is evaluated at an intermediate lag time of $\tau=50$ for the self-driven rod in free space. Circles denote simulation data, and solid lines connect the data points as guides to the eye.}
  \label{vanHove_SI}
\end{figure*}
%=========================================================================

%=========================================================================
\begin{figure*}
    \includegraphics[width=10cm]{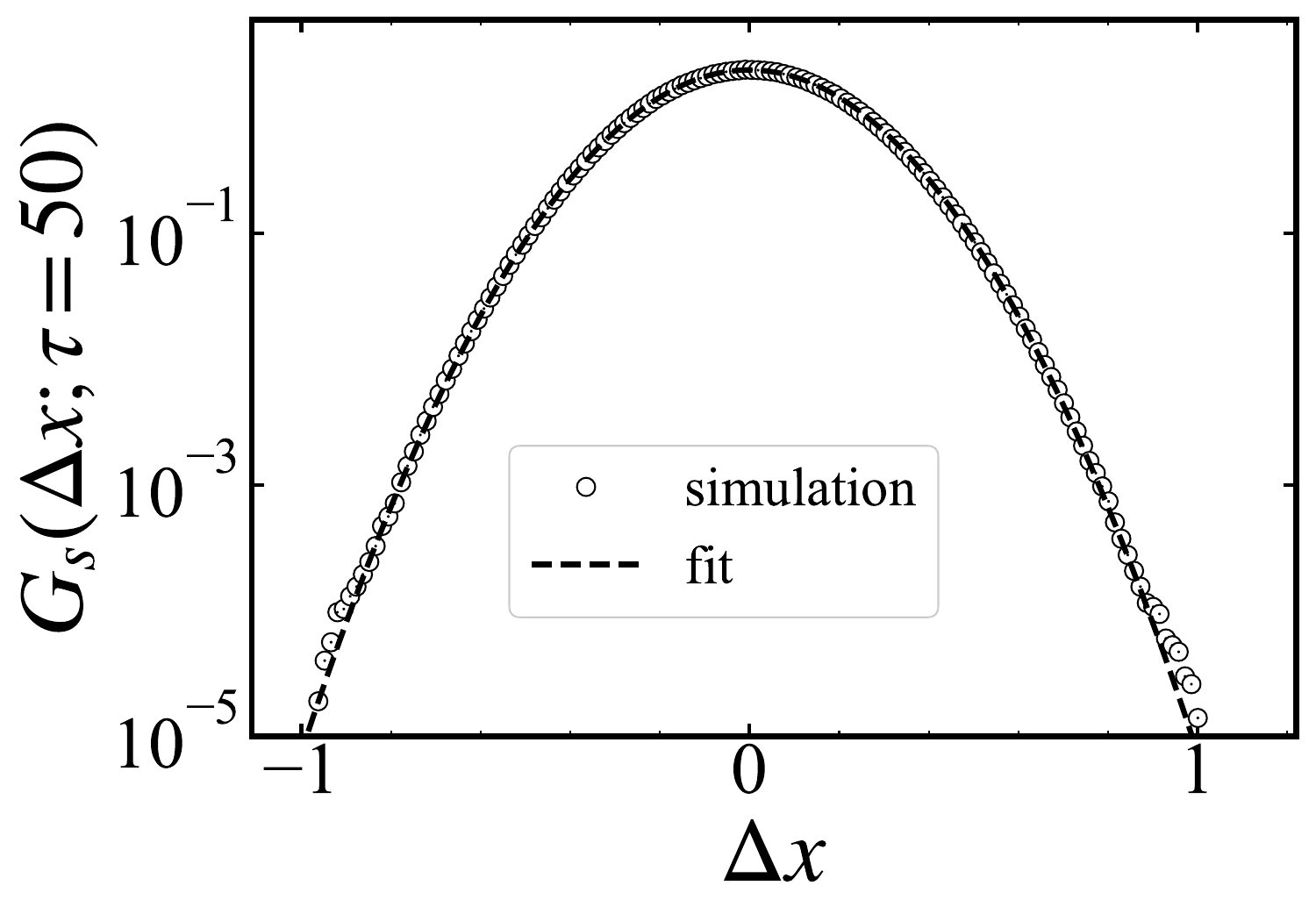}
  \caption{Analytical Gaussian fit of the self-part of the van Hove correlation function, $G_s(\Delta x, \tau)$, for the unconfined passive nanorod. The spatial distribution is evaluated at a lag time of $\tau = 50$ in free space. Black circles denote the simulation data, and the dashed line represents the corresponding analytical Gaussian profile.}
  \label{vanHove_gaussian_fit}
\end{figure*}
%=========================================================================

%=========================================================================
\begin{figure}
    \includegraphics[width=\textwidth]{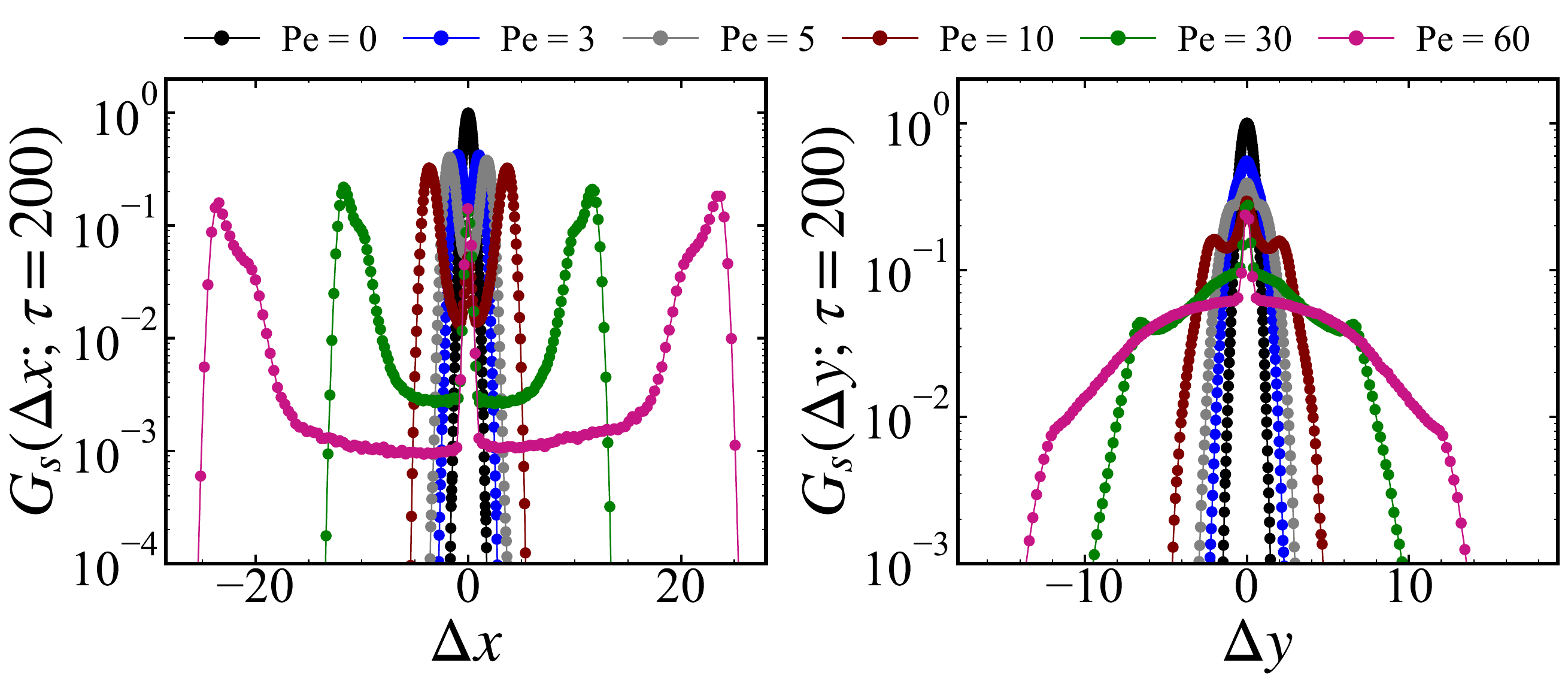}
  \caption{van-Hove function of the active rod in a SC channel at lag time $\tau = 200$. The left panel shows the VHCF of the rod along x-directions, and right panel corresponds to the y-direction. The solid lines are guiding the data points.}
  \label{vanHove_lt_200_SCC}
\end{figure}
%=========================================================================

%=========================================================================
\begin{figure}
    \includegraphics[width=\textwidth]{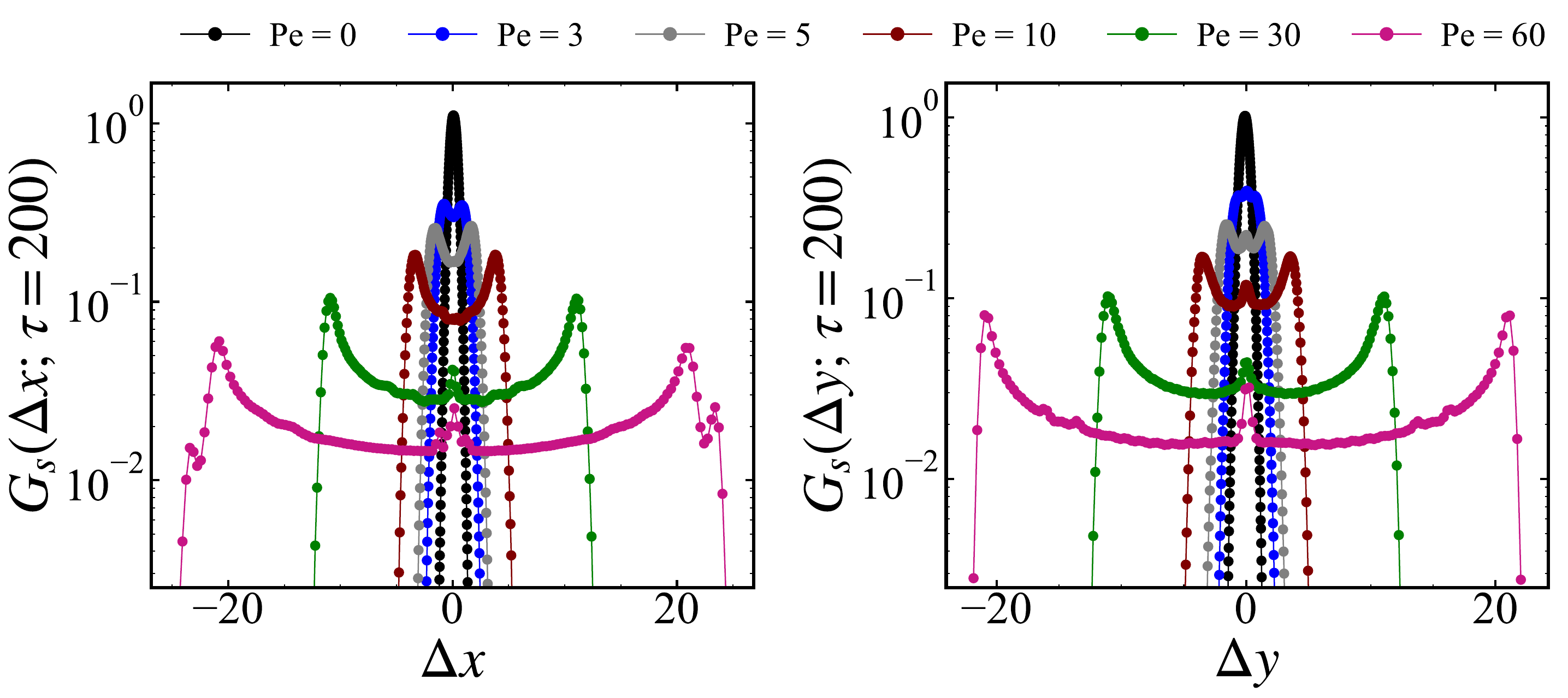}
  \caption{van-Hove function of the active rod in a CC channel at lag time $\tau = 200$ with behavior along both x and y-direction.}
  \label{vanHove_lt_200_CCC}
\end{figure}
%=========================================================================

%%%REFERENCES%%%
%\bibliography{rsc} %You need to replace "rsc" on this line with the name of your .bib file
%\bibliographystyle{rsc} %the RSC's .bst file

\end{document}